\documentclass[twocolumn,preprintnumbers]{revtex4}
\usepackage{amssymb}
\usepackage{amsmath}
\usepackage{graphicx}
\usepackage{graphics}
\usepackage{dcolumn}
\usepackage{bm}

\begin{document}

\title{Lower bound for the population of hyperfine component $\mu =0$
particles in the ground state of spin-1 condensates}

short title: Lower bound for the hyperfine population
 of spin-1 condensates

\author{C. G. Bao}
\affiliation{The State Key Laboratory of Optoelectronic Materials and \\
Technologies School of Physics and Engineering, Sun Yat-Sen University,
Guangzhou, P. R. China}
\email{stsbcg@mail.sysu.edu.cn}

\pacs{03.75.Mn,03.75.Kk}

\begin{abstract}
An analytical expression for the lower bound of the average number of
hyperfine component $\mu =0$ particles in the ground state of spin-1
condensates (denoted as $\overset{\_\_}{\rho _{0}}$) under a magnetic field
has been derived. In the derivation the total magnetization $M$ is kept
rigorously conserved. Numerical examples are given to show the applicability
of the analytical expression. It was found that, in a broad domain of
parameters specified in the paper, the lower bound is very close to the
actual $\overset{\_\_}{\rho _{0}}$. Thereby, in this domain, $\overset{\_\_}{%
\rho _{0}}$ can be directly evaluated simply by using the analytical
expression.
\end{abstract}

\maketitle

\bigskip
\bigskip

The spinor condensates, as tunable systems with active spin-degrees of
freedom, are rich in physics and promising in application. Since the
pioneering experiment on spin-1 condensates \cite{ste} the study of these
systems becomes a hot topic. In the study, an important observable is the
probability density of the particles in a given hyperfine-component $\mu
=\pm 1$, or $0$. These quantities are popularly measured in various
experiments and are a key to relate experimental results to theories.\cite%
{ste,dms,hsc,msc,tku} Recently, Tasaki has derived an inequality for the
lower bound of the average number of $\mu =0$ particles $\overset{\_\_}{N_{0}%
}$ in the ground state (g.s.) of the spin-1 condensates under a magnetic
field $B$.\cite{tasa13} In his derivation the term $\langle \Phi _{GS},\hat{V%
}\Phi _{GS}\rangle $ (where $\Phi _{GS}$ is the g.s. wave function and $\hat{%
V}$ is the total interaction) has been considered as zero. Since this term
is not a small term but an important term, the constraint given by his
inequality is very loose. In particular, when the parameters of the system
are given in a broad domain frequently accessed in related experiments, the
lower bound of $\overset{\_\_}{N_{0}}$ appears as a negative value (see
below). Hence, in order to have an applicable lower bound, the inequality by
Tasaki should be substantially improved.

In this paper the inequality has been re-derived by taking the missing and
important term $\langle \Phi _{GS},\hat{V}\Phi _{GS}\rangle $ back into
account. In this way, as shown below, a much higher lower bound together
with a upper bound for $\overset{\_\_}{N_{0}}$ can be obtained. It is found
that, in some cases, the lower bound is very close to the upper bound, and
thereby $\overset{\_\_}{N_{0}}$ can be directly evaluated. Besides, the
inequality by Tasaki is only for the case with the total magnetization $M=0$%
. However, under a magnetic field $B$, $M$\ is a good quantum number and its
magnitude depends on how the condensate is experimentally prepared. Thus the
g.s. is not necessary to have $M=0$. Therefore, an arbitrary $M\geq 0$ is
considered in the follows.

When $B\neq 0$, due to the conservation of $M$, the linear Zeeman energy is
a constant and hence irrelevant. Thus the Hamiltonian can be written as\cite%
{tasa13}
\begin{equation}
\hat{H}=\hat{H}_{0}+\hat{V}-q\hat{N}_{0}  \label{1}
\end{equation}%
where $\hat{H}_{0}=\sum_{i}(-\frac{\hbar ^{2}}{2m}\bigtriangledown
_{i}^{2}+U(r_{i}))$ includes the kinetic and trap energies, $\hat{V}=\hat{V}%
_{0}+\hat{V}_{2}$, $\hat{V}_{0}=c_{0}\sum_{i<j}\delta (\mathbf{r}_{i}\mathbf{%
-r}_{j})$ and $\hat{V}_{2}=c_{2}\sum_{i<j}\delta (\mathbf{r}_{i}\mathbf{-r}%
_{j})\ \mathbf{f}_{i}\mathbf{.f}_{j}$, where $\mathbf{f}_{i}$ is the
spin-operator of the i-th particle. The third term arises from the quadratic
Zeeman energy where $\hat{N}_{0}$ is the operator for the number of $\mu =0$
particles.

Let $S$ be the total spin of the $N$ spin-1 atoms, and $M$ is the
Z-component of $S$. Let $\vartheta _{SM}^{[N]}$ be the total spin-state with
good quantum numbers $S$ and $M$. It has been proved that $\vartheta
_{SM}^{[N]}$ is unique (i.e., there is only one $\vartheta _{SM}^{[N]}$ for
a pair of $S$ and $M$), $N-S$ must be even, and $\{\vartheta _{SM}^{[N]}\}$
is a complete set for all symmetric total spin-state of spin-1 atoms.\cite%
{ka} Thus they can be used as basis functions in the follows.\cite{ka}

Due to the third term in $\hat{H}$, different $\vartheta _{SM}^{[N]}$
distinct in $S$\ are mixed up in the g.s., as
\begin{equation}
\Phi _{GS}=F(\mathbf{r}_{1},\cdot \cdot \cdot \mathbf{r}_{N})\sum_{S}C_{S}%
\vartheta _{SM}^{[N]}\ \ \ \   \label{2}
\end{equation}
where the function $F$ for the spatial degrees of freedom and the
set of coefficients $\{C_{S}\}$ are unknown. For any given state
$\Psi \neq \Phi _{GS}$. Obviously,
\begin{equation}
\langle \Phi _{GS},\hat{H}\Phi _{GS}\rangle \leq \langle \Psi ,\hat{H}\Psi
\rangle  \label{3}
\end{equation}%
This equation is the base for the following derivation.\cite{tasa13}

Let $\Psi \equiv F\cdot |M,N-M,0\rangle $, where $|N_{1},N_{0},N_{-1}\rangle
$\ denotes a Fock-spin-state with $N_{\mu }$ particles in $\mu -$component.
Since $\Phi _{GS}$ and $\Psi $\ have exactly the same spatial wave function,
it is straight forward to prove that $\langle \Phi _{GS},(\hat{H}_{0}+\hat{V}%
_{0})\Phi _{GS}\rangle =\langle \Psi ,(\hat{H}_{0}+\hat{V}_{0})\Psi \rangle $%
. Let $\int d\mathbf{R}\delta (\mathbf{r}_{i}\mathbf{-r}_{j})|F|^{2}\equiv X$%
, where the integration covers all the spatial degrees of freedom. Note that
$X$\ does not depend on $i$ and $j$\ due to the symmetry inherent in $F$.
Making use of the formulae $\sum_{i<j}\mathbf{f}_{i}\mathbf{.f}_{j}=\frac{1}{%
2}\hat{S}^{2}-N$, $\langle M,N-M,0|\hat{S}^{2}|M,N-M,0\rangle =(M+1)(2N-M)$
(say, refer to eq.(A5) of \cite{ml}), and $\langle \Psi ,\hat{N}_{0}\Psi
\rangle =N-M$,\ eq.(\ref{3}) becomes
\begin{eqnarray}
&&\langle \Phi _{GS},\hat{\rho}_{0}\Phi _{GS}\rangle \equiv \frac{1}{N}\langle \Phi _{GS},%
\hat{N}_{0}\Phi _{GS}\rangle   \\ &&\geq \frac{N-M}{N}
+\frac{c_{2}X}{2qN}[\sum_{S}C_{S}^{2}S(S+1)-(2N-M)(M+1)] \nonumber
\end{eqnarray}
where $\langle \Phi _{GS},\hat{\rho}_{0}\Phi _{GS}\rangle \equiv \overset{%
\_\_}{\rho _{0}}$ is the probability density of $\mu =0$ component in the
g.s..

Since $N\geq S\geq M$, disregarding how $C_{S}$ is,\ we have $N(N+1)\geq
\Sigma _{S}C_{S}^{2}S(S+1)\geq M(M+1)$. Thus, for $c_{2}>0$ (say, Na atoms),
$\Sigma _{S}C_{S}^{2}S(S+1)$ can be safely replaced by its lower limit $%
M(M+1)$, and we have
\begin{equation}
\overset{\_\_}{\rho _{0}}\geq \frac{N-M}{N}[1-c_{2}\frac{(M+1)}{q}%
X],~~(c_{2}>0)  \label{5}
\end{equation}%
Whereas for $c_{2}<0$ (say, Rb atoms), $\Sigma _{S}C_{S}^{2}S(S+1)$ can be
safely replaced by its upper limit $N(N+1)$, and we have
\begin{equation}
\overset{\_\_}{\rho _{0}}\geq \frac{N-M}{N}[1-|c_{2}|\frac{N-M-1}{2q}%
X],~~(c_{2}<0)  \label{6}
\end{equation}

The right sides of \ref{5} and \ref{6} are just the lower bounds for $%
\overset{\_\_}{\rho _{0}}$\ denoted as $(\overset{\_\_}{\rho _{0}})_{low}$,
where $X$ can be estimated in various ways. For instance, when one assume
that all the particles have the same spatial wave function $\phi (\mathbf{r}%
) $, then $X=\int |\phi |^{4}d\mathbf{r}$. Where, $\phi $ can be obtained by
minimizing the Hamiltonian under the assumption $\Phi _{GS}=\Pi _{i}\phi (%
\mathbf{r}_{i})\sum_{S}C_{S}\vartheta _{SM}^{[N]}\ $.\cite{bao3}

If the Thomas-Fermi (TF) approximation is further applied (i.e., neglecting
the kinetic energy and the spin-dependent force), the
resulting single particle wave function $\phi _{TF}(\mathbf{r})$ can be
easily obtained. Accordingly, we have $X_{TF}\equiv \int |\phi _{TF}|^{4}d%
\mathbf{r}$. Since $|c_{2}|$ is two order smaller than $c_{0}$, the effect of $|c_{2}|$ on the spatial wave functions is very small (say, refer to the numerical results given in Fig.2a of \cite{bao1}). Moreover, the neglect of the kinetic energy will lead to a more compact distribution of the wave function and therefore a larger $X_{TF}$. Therefore, it is a reasonable approximation to assume that, due to the combined effect of neglecting the kinetic energy and the spin-dependent force, we have $X_{TF} \geq X$. Thus, since the $X$ in eq.(5) or (6) is multiplied by a negative value, the inequality remains hold when $X$ is  replaced by $X_{TF}$.

When $c_{2}=0$ and the trap $U(r_{i})=\frac{1}{2}m\omega ^{2}r_{i}^{2}$, $%
X_{TF}$ has a very simple form, it reads $X_{TF}=0.3067(N\overset{\_\_}{c_{0}%
})^{-3/5}\lambda ^{-3}$, where $\overset{\_\_}{c_{0}}$\ is the value of $%
c_{0}$ when $\hbar \omega $ and $\lambda \equiv \sqrt{\hbar /(m\omega )}$
are used as units for energy and length, respectively (i.e., $c_{0}=\overset{%
\_\_}{c_{0}}\hslash \omega \lambda ^{3}$). With this in mind, eq.(5) and (6)
can be rewritten in a more concise way as
\begin{equation}
\overset{\_\_}{\rho _{0}}\geq \frac{N-M}{N}[1-\frac{|c_{2}|}{q}K_{M}X_{TF}]
\label{7}
\end{equation}
where $K_{M}=M+1$ (if $c_{2}>0$) or $(N-M-1)/2$ (if $c_{2}<0$).

For $M=0$, eq.\ref{7} can be compared with the inequality by Tasaki. The
parameters used in this paper and in \cite{tasa13} are related as $%
c_{0}=(2g_{2}+g_{0})/3$ and $c_{2}=(g_{2}-g_{0})/3$. Then, the inequality of
Tasaki is \cite{tasa13}%
\begin{equation}
\overset{\_\_}{\rho _{0}}\geq 1-c_{0}\frac{N}{2qV_{eff}},  \label{8}
\end{equation}
where $V_{eff}$\ is named the effective volume, and is so defined that $%
1/V_{eff}$ is the smallest constant $\geq |\varphi _{0}(\mathbf{r})|^{2}$
for any $\mathbf{r}$, and $\varphi _{0}(\mathbf{r})$\ is the single-particle
ground state of $\hat{H}_{0}$. When the trap is $\frac{1}{2}m\omega
^{2}r^{2} $, $V_{eff}=\pi ^{3/2}\lambda ^{3}$. Using the units $\hbar \omega
$, $\lambda $, and $\sec $, $c_{2}=\overset{\_\_}{c_{2}}\hslash \omega
\lambda ^{3}$, $\omega =\overset{\_}{\omega }\sec ^{-1}$, $q=\overset{\_}{q}%
\hbar \omega $. For Na (Rb), the related dimensionless quantities are $%
\overset{\_\_}{c_{0}}=6.77\times 10^{-4}\sqrt{\overset{\_\_}{\omega }}\ $($%
2.49\times 10^{-3}\sqrt{\overset{\_\_}{\omega }}$),\ $\overset{\_\_}{c_{2}}%
=2.12\times 10^{-5}\sqrt{\overset{\_\_}{\omega }}\ $($-1.16\times 10^{-5}%
\sqrt{\overset{\_\_}{\omega }}$). Let $\gamma $\ be the ratio of the second
terms of (7) over the one of (8). For Na (Rb) with $M=0$, $\gamma
=8.53/[N^{8/5}(\overset{\_}{\omega })^{3/10}]$ ($0.29/[N^{3/5}(\overset{\_}{%
\omega })^{3/10}]$). Since $N$\ and $\overset{\_}{\omega }$\ are usually
large, $\gamma $ is usually very small. It implies that the lower bound
given by (7) is much higher.

Let $B=\overset{\_}{B}$ (Gauss), then $\overset{\_}{q}=1745\overset{\_}{B}^{2}/%
\overset{\_}{\omega }$ ($452\overset{\_}{B}^{2}/\overset{\_}{\omega }$) for
Na (Rb). To relate the lower bound directly with $\overset{\_}{B}$, eq.(7)
can be rewritten as
\begin{equation}
(\overset{\_\_}{\rho _{0}})_{low}=\frac{N-M}{N}[1-Y_{M}(\overset{\_}{\omega }%
^{2}/N)^{3/5}/\overset{\_}{B}^{2}]  \label{9}
\end{equation}
where $Y_{M}=2.97\times 10^{-7}(M+1)$ ($1.44\times 10^{-7}(N-M-1)$)
for Na
(Rb). Obviously, when $M$\ is conserved, the upper bound $(\overset{\_\_}{%
\rho _{0}})_{up}$ is just $(N-M)/N$. Therefore, for Na (Rb), when $Y_{M}$
and $\overset{\_}{\omega }^{2}/N$ are small, the lower bound will be higher
and close to the upper bound. Otherwise, the lower bound might be too low
and becomes meaningless. In any case, the lower bound will be higher when $%
\overset{\_}{B}$ is larger, and will become meaningless when $\overset{\_}{B}%
\rightarrow 0$. Numerical examples are given in Table \ref{table.1}.

\begin{table}[pb]
\caption{Examples of the lower bound $(\overset{\_\_}{\rho
_{0}})_{low}$\
from eq.(9) with $N=10^{5}$. $u$ denotes the upper bound, namely, $u\equiv (\overset{\_\_}{\rho _{0}}%
)_{up}=(N-M)/N$ (specifically, $u=1$ when $M=0$, and $u=10^{-3}$ when $M=N-100$). $(%
\overset{\_\_}{\rho _{0}})_{low}$\ from eq.(8) are given in the
parentheses (only for $M=0$). The upper part of the table is for
$\overset{\_}{B}=0.1$ (Gauss), while the lower part
is  $\overset{\_}{B}=1$ (Gauss).}%
\label{table.1}
\begin{center}%
\begin{tabular}
[c]{ccccc}%
$\overset{\_}{B}=0.1$& &$\overset{\_}{\omega }=30$  & &$\overset{\_}{%
\omega }=300$  \\
Na, M=0 & &$u-1.8\times 10^{-6}$  & & $u-2.8\times
10^{-5}$  \\
Na, M=0 & &(-56.2) & &(-1809.2)  \\
Rb, M=0 & &$u-8.5\times 10^{-2}$  & &$u-1.348$  \\
Rb, M=0 & &(-811.9)  & &(-25700) \\
Na, M=N-100 & &$u-1.8\times 10^{-4}$ & & $u-2.8\times
10^{-3}$ \\
Rb, M=N-100 & &$u-8.0\times 10^{-8}$& & $u-1.3\times 10^{-6}$ \\ \\
$\overset{\_}{B}=1$& &$\overset{\_}{\omega }=30$ & &$\overset{\_}{%
\omega }=300$ \\
Na, M=0  & &$u-1.8\times 10^{-8}$  & &$u-2.8\times 10^{-7}$ \\
Na, M=0  & &(0.43) & &(-17.1) \\
Rb, M=0 & & $u-8.5\times 10^{-4} $ & & $u-1.348\times 10^{-2}$ \\
Rb, M=0  & &(-7.1)  & &(-256) \\
Na, M=N-100  & &$u-1.8\times 10^{-6}$ & & $u-2.8\times 10^{-5}$ \\
Rb, M=N-100 & &$u-8.0\times 10^{-10}$ & & $u-1.3\times 10^{-8}$%
\end{tabular}
\end{center}
\end{table}

Table \ref{table.1} demonstrates that, for Na (Rb), when $M/N$
($(N-M)/N$) is small the lower bound $(\overset{\_\_}{\rho
_{0}})_{low}$\ is very close to its upper
bound, and therefore is close to the actual density $\overset{\_\_}{\rho _{0}%
}$. In particular, it is even closer when $\omega $ is small and $B$ is
larger. However, for Na (Rb), when $M/N$ ($(N-M)/N$) is close to 1, the
constraint provided by eq.(9) is loose. It is even worse when $\omega $ is
larger and $B$ is smaller. (say, in the column with $\overset{\_}{\omega }%
=300$ and $\overset{\_}{B}=0.1$, both the values $u-1.348$\ and
$u-2.8\times 10^{-3}$\ are negative). In this case, the lower bound
is completely meaningless. For all the cases under consideration,
the lower bounds given by eq.(8) are negative as shown by the values
inside the parentheses.

When $q\rightarrow 0$, all the above inequalities do not work. In this case,
it is suggested that the perturbation theory could be used to evaluate $%
\overset{\_\_}{\rho _{0}}$ (this is beyond the scope of this paper). In
particular, when $B=0$, it has been derived in \cite{bao2} that $\overset{%
\_\_}{\rho _{0}}=(N-M)/(N(2M+3))$ ($c_{2}>0$), or $=(N-M)(N+M)/[N(2N-1)]$ ($%
c_{2}<0$). These two formulae is a generalization of the \textit{theorem 1}
in \cite{tasa13}.

In conclusion, although the derivation of the inequality by Tasaki is
rigorous, the resulting lower bound is too low to be meaningful. On the other hand, although
approximations have been used in this paper (namely, the M-conserved SMA and
the T-F approximation), by recovering the important term $\langle \Phi _{GS},%
\hat{V}\Phi _{GS}\rangle $ which has been omitted in \cite{tasa13}, the
lower bound given in this paper is remarkably higher. In particular,

(i) The inequality has been generalized for the case with an arbitrary $%
M\geq 0$. The generalization to the case with a negative $M$ is straight forward,

(ii) The constraint is now species-dependent. Since the g.s. of the Rb
condensate is greatly different from the Na condensate, this dependence is
reasonable.

(iii) In a broad domain of parameters frequently accessed in experiments,
the new inequality is applicable. In particular, when $Y_{M}(\overset{\_}{%
\omega }^{2}/N)^{3/5}/\overset{\_}{B}^{2}$ is sufficiently small, the
resulting $(\overset{\_\_}{\rho _{0}})_{low}$ is very close to $(\overset{%
\_\_}{\rho _{0}})_{up}$ (as shown in Table 1), and therefore a direct evaluation of $\overset{\_\_}%
{\rho _{0}}$ can be achieved. Otherwise, $(\overset{\_\_}{\rho _{0}})_{low}$
will deviate remarkably from $(\overset{\_\_}{\rho _{0}})_{up}$. In this
case $\overset{\_\_}{\rho _{0}}$ can not be evaluated accurately.


\begin{thebibliography}{99}
\bibitem{bao3} The single spatial mode approximation (SMA) mentioned here is
not identical to the one widely used within the framework of the mean-field
theory. In the latter, $M$ is not conserved, but its average is fixed.
Whereas in the former $M$ is exactly conserved as it should be. Since the
two SMA are different in nature, their validities can not be judged by the
same criterion. An example of the $M-$conserved SMA is referred to the ref.%
\cite{bao2}.
\end{thebibliography}
\end{document}